# Integrating Positionality Statements in Empirical Software Engineering Research

Breno Felix de Sousa
*Federal University of Pernambuco*
Recife, Brazil
bfs7@cin.ufpe.br

Ronnie de Souza Santos
*University of Calgary*
Calgary, Canada
ronnie.desouzasantos@ucalgary.ca

Kiev Gama
*Federal University of Pernambuco*
Recife, Brazil
kiev@cin.ufpe.br

*Abstract*—*Context*. Positionality statements are a reflective practice that is a well-established practice in fields such as social sciences, where they enhance transparency, reflexivity, and ethical integrity by acknowledging how researchers' identities, experiences, and perspectives may shape their work. *Goal*. This study aimed to investigate the understanding, usage, and potential value of positionality statements in software engineering (SE) research, particularly in studies focused on diversity and inclusion (D&I). *Method*. We conducted a qualitative survey targeting authors of D&I-focused studies in SE to explore their perspectives and practices regarding positionality. Through purposive sampling, we collected responses from 21 participants, which were analyzed using thematic analysis to identify how positionality is currently understood and applied. *Findings*. Our findings reveal that SE researchers often view positionality statements as a method for self-reflection, contextual awareness, and bias reduction, though practices vary widely. While some participants explicitly integrate positionality statements into their research, most apply these concepts implicitly. Challenges, such as double-anonymity requirements and the perception of objectivity in SE, also limit the adoption of positionality. *Discussions*. Our findings highlight an opportunity for SE to adopt and adapt positionality statements to reflect the field's intersection of technical and human considerations. By incorporating structured positionality practices, SE research could enhance inclusivity, ethical rigor, and transparency, moving closer to the standards established in more mature disciplines. *Conclusion*. Although SE research increasingly addresses complex social and human-centered issues, positionality statements have yet to become common practice in the field.

## I. INTRODUCTION

The concept of reflexivity in qualitative research involves a multifaceted process of self-examination where researchers critically assess how their personal background and the research context might shape their work [1], [2]. This continuous process acknowledges subjectivity by making clear that researchers are not objective observers but active participants whose viewpoints influence the research [3]. While reflexivity encompasses a range of practices, one increasingly popular method is the inclusion of positionality statements [3], [4]. A positionality statement, also known as a reflexivity statement [4], is a practice in which researchers explicitly disclose their positions, beliefs, and personal experiences that may influence their study—whether in methodological decisions, interpretation of findings, or addressing threats to validity [4]–[9]. These statements are common and well-established across various fields. For instance, social sciences have a longstanding tradition of defining and applying positionality statements, particularly in qualitative methodologies [10], [11]. Furthermore, interdisciplinary fields in computing, such as Computer-Supported Cooperative Work (CSCW) and Computer-Human Interaction (CHI), have also been actively discussing the role of positionality in their studies [6], [12]–[14]. However, only recently there has been a major publication raising concerns on reflexivity, by stating that software engineering researchers seldom reflecting on their assumptions and biases [].

By disclosing their positionality, researchers acknowledge and detail aspects of their identities and experiences that could shape their approach to research, potentially influencing how they conduct, interpret, and communicate their study's findings [15]. It enables them to move beyond the common technical threats to validity mainly associated with methods and consider more nuanced ways in which personal biases and identity factors (e.g., gender, race, socioeconomic status, ethnicity, sexuality, geographic location, and other intersectional aspects) inform their perspectives and interactions throughout the research process [16]–[18]. Such reflection allows researchers to consider potential influences on their study and to discuss the employed strategies to mitigate these effects.

Positionality statements are a standard practice in social sciences – with many journals explicitly asking authors for such statements [4], [8] – and increasingly common in SE-related fields, such as Computer-Supported Cooperative Work (CSCW) and Human–computer interaction (HCI) [14], and broader areas such as Computer Science Education [19]. However, they are rarely used in Software Engineering (SE) qualitative research. This is particularly interesting, especially considering Software Engineering's reliance on qualitative methods, which, in other fields, frequently involve explicit disclosures of positionality. In this context, this study explores the current state of positionality statement usage in Software Engineering, motivated by the potential of these statements to enrich understanding, mitigate biases, and foster a reflective research culture, particularly in studies addressing diversity, inclusion, and other social aspects of the field. Specifically, we seek to answer the following research question **(RQ):** *How are positionality statements used in Software Engineering studies?*

To address this RQ, we analyzed the articles gathered in the most relevant systematic literature review (SLR) to date on Diversity and Inclusion (D&I) in software engineering (SE) [20]. We looked for occurrences of positionality statements in that corpus and surveyed authors about their perspective. We chose that SLR as the starting point for analyzing positionality statements because D&I research inherently engages with issues of identity, bias, and marginalization, making positionality particularly relevant. In addition, an existing SLR ensures methodological rigor and a representative sample, providing a solid foundation for identifying whether and how SE researchers acknowledge their positionality.

## II. BACKGROUND

While positionality statements are used to enhance research transparency, they also support social justice and ethical research practices [11]. By acknowledging their own positions on the topic under investigation, researchers demonstrate an awareness of how their status and perspectives may influence study participants and the research process itself [8]. For example, when studying vulnerable communities, researchers must consider how their socioeconomic or cultural backgrounds might shape the relationships they build with these communities and the interpretation of their experiences [21].

In social sciences, where the focus often includes human behavior, social relationships, and power structures, it is essential to recognize that researchers are not fully neutral or objective agents, as their perspectives are influenced by their own social positions [17], [22]. Acknowledging privilege or disadvantage, therefore, allows researchers to reflect on how their positionality may impact their interactions and findings. Bringing this practice into SE, where research on diversity and inclusion is increasing, we observe the need for collective reflection on the responsibility of the research community to engage thoughtfully with these themes [23].

In this context, when adopting methodological strategies involving in-person interactions, such as interviews and observations, building trust with communities such as LGBTQIA+ software professionals, ethnically underrepresented SE students, and practitioners facing discrimination, among others, requires researchers to demonstrate that they can be trusted as allies, even if they are not part of the minority group focused on in the study [23], [24]. In these cases, positionality statements can serve as an important practice, helping researchers convey their awareness and commitment to understanding the perspectives of those they study while also being transparent about their own.

In general, explicit positionality statements in research allow researchers to disclose and address potential normative biases, build trust, and avoid imposing dominant perspectives that could harm underrepresented groups [25], [26]. Even in studies involving populations not considered underrepresented, acknowledging positionality can improve the understanding of sociocultural dynamics within the field and foster a more critical perspective in engineering [5]. In other words, since scientific research is conducted by people, positionality provides safeguards against potential biases inherent in human analysis. Awareness of positionality encourages a more critical assessment on study validity and reliability [4], [8], [9], [26].

While the importance of positionality statements is widely acknowledged, there has been also criticism. Due to becoming a standard practice in Social Sciences, where many journals are asking for an explicity positionality statement [4], [8], there is a growing concern on this becoming rather a shallow "checklist" that does not necessarily lead to reflexivity [3], [8].

The way readers see the author's positionality (i.e., perceived positionality) may also affect how the work is received or understood [27]. This can have consequences also in peer review. Despite the benefits of enhancing transparency from authors, it has also the potential to increase bias from reviewers in the peer review process [14].

Positionality is a topic taking part of recent discussions of Special Interest Groups in the CHI and CSCW communities [28], [29]. In an HCI literature review about privacy and marginalized groups [30], the authors identified only 23% of articles in the 88 papers dataset having. Positionality statements were often brief, serving primarily to convey the authors' identities—whether marginalized or not—and most papers did not discuss how these positionalities might inform or impact the research. Thus, reinforcing concerns raised by some scholars [4], [14]. While other CS-related fields doing qualitative studies are advancing in such discussions, reflexivity or positionality statements were not part of popular methodology checklists of SE qualitative methods borrowed from social sciences, such as ethnography [31], [32]. However, they are on the SE radar. A reference on reflexivity is cited in the ACM SIGSoft empirical standards for qualitative surveys [33]. Also, the importance of reflexivity was recently highlighted as essential for maintaining transparency, enhancing the trustworthiness of findings, and ensuring ethical interactions between researchers and participants [34]. Authors underscored the scarcity of such practice in SE.

## III. METHODOLOGY

In this study, we adopted a mixed study that employs content analysis of an existing Systematic Literature Review [20] and a qualitative survey approach to explore the role of positionality statements in software engineering research. This approach was well-suited to our research question, as it draws on participants' perspectives and experiences, allowing us to capture a broad range of views on a specific topic.

### A. Positionality statements search

In the reference SLR [20] a set of 131 studies about diversity in Software Engineering was compiled, covering up to May 2020. Our goal using that dataset was to identify how researchers in the SE field have documented their positionality and in what contexts such statements are inserted. We downloaded the articles listed in this dataset to systematically search for positionality statements.

Two authors performed searches in those articles using the pdfgrep tool to look for the terms "reflexivity", "positionality", "position statement" and "bias". The resulting set respectively consisted of 0, 0, 2 and 85 files with occurrences of those term, leading to 87 articles in total. The first three terms would consist of direct usage of positionality, while "bias" occurrences lead to a thorough reading of articles in order to infer context that would suggest positionality or authors explicitly mentioning they belong to underrepresented groups.

## B. Survey

**Sampling and Participants** For this study, our population consisted of 285 valid [1] author emails extracted from the 131 studies focused on D&I in software engineering, who were listed in a recent systematic literature review on the theme [20]. We chose to sample from this group because positionality statements are closely tied to human factors, particularly those related to equity, diversity, and inclusion. Our sampling strategy can be characterized as purposive sampling [35], which involves deliberately selecting participants who are knowledgeable about a specific topic.

In this study, we report findings from the first cycle of data collection, i.e., the analysis of responses received so far, totaling 21 authors who answered the survey. The demographics of our sample are as follows: 14.3% (N=3) identified as Lesbian, Gay, Bisexual, Transgender, Queer, Intersex, Asexual, or other (LGBTQIA+); 38.3% (N=8) identified as women; and 47.6% (N=10) identified as belonging to non-underrepresented groups. For clarity, these percentages were rounded to one decimal place. Among the participants, intersecting identities included Women LGBTQIA+ 4.8% (N=1), Women Black 4.8% (N=1), Women Latina 4.8% (N=1), and LGBTQIA+ Neurodivergent 4.8% (N=1). Each of these intersections represents 4.8% of the sample, illustrating a diverse range of perspectives within the participant group. The diversity of our sample offers a range of perspectives that can help us examine positionality in software engineering through a diversity lens similar to those used in social sciences and related fields.

**Data Collection.** Data were collected using a semi-structured questionnaire crafted to capture participants' experiences and views regarding positionality statements in their studies. The list of authors extracted from the SLR served as the basis for sending invitation emails for the study. When contacting these authors, we explained that we identified their work through an SLR focused on perceived diversity in software engineering. We also highlighted that we were conducting a survey on the use of positionality statements in research in the field and that we would like to understand how these researchers perceive and use these statements in their studies. The survey included seven open-ended questions as detailed in Table I. Data collection began on April 24, 2024.

**Data Analysis.** We followed well-established guidelines to conduct our data analysis based on the method of thematic analysis [36], [37]. Based on this method, we analyzed participants' responses by following these steps:

1) **Familiarization**: Carefully reading through all responses to gain an overall understanding of the content.

---
[1] Out of the messages sent to the initial extraction of 353 emails, 68 failed

TABLE I
SURVEY QUESTIONS

| |
|---|
| 1. Do you consent to participate in this research? |
| 2. Do you conduct research on Diversity and Inclusion (D&I) in Software Engineering? |
| 3. Have you published any articles in the Empirical SE field that address minorities (e.g., women, LGBTQIA+, Black/Brown individuals, Indigenous people)? |
| 4. Are you a member of any underrepresented group in computing (e.g., LGBTQIA+, women, Black individuals)? If so, which one? |
| 5. What do you understand by "positionality statement"? |
| 6. If you or any co-author has explicitly used a positionality statement in your D&I-related publications, could you briefly describe how it was done? |
| 7. In your D&I research, how do you avoid personal biases stemming from your perspective? |

2) **Initial Coding**: Identifying key codes directly from the data to capture relevant ideas and patterns.
3) **Theme Development**: Grouping related codes to define and label overarching themes that represent the data.
4) **Theme Review**: Refining themes to ensure clarity, relevance, and distinctiveness.
5) **Report Production**: Compiling the results into a coherent and meaningful narrative to present the findings.

This process was supported by using MaxQDA https://www.maxqda.com/quantitative-text-analysis, which facilitated the organization, categorization, and interpretation of data through our thematic analysis[2].

**Ethical Considerations.** This study adhered to ethical standards for research involving human participants and received approval from the first author's university ethics committee. Before starting the survey, participants were informed about the study's purpose, anonymous voluntary participation, and confidentiality and provided informed consent. Survey data were securely stored and accessible only to the research team.

## IV. FINDINGS

### A. Positionality statements

There were no occurrences of "reflexivity" or "positionality" in the dataset, two occurrences of "position statement", and 85 occurrences of "bias". While most of them consisted of gender bias studies, after analyzing them, only four indirectly discussed positionality. We discuss them below and indicate their category (e.g., case study, ethnography) in the SLR [20].

In the two occurrences of "position statement", one just had a citation with that term in the tile and the other [38] (mixed study) explicitly had "Position Statement" subsection in the method. They stated that *"the research team includes Black women with strong interests in disrupting the dominant sociocultural norms in computing. We represent over 18 years of conducting research that aims to promote a more complex narrative of the ways Black women's bodies move through the world, particularly in computing."* It indicates the authors' standpoint of those who are immersed in the problem and have practical knowledge of the topic due to their personal and professional experiences as Black women in computing. However, the biases of such close connection the subject were not discussed further.

---
[2] The data supporting the findings of this study are available on Figshare https://figshare.com/s/121111560eae59a64fa3

When manually analyzing the articles looking for statements that would indirectly suggest positionality, we found 4 articles [39]–[42]. In [39] there was the explicit mention of authors' gender in the method section *"Both (female) researchers facilitated the workshop"*, with a brief discussion to minimize interview bias by debriefing checking for similar or discrepant interpretations. The authors of [40] (survey study) were investigating the perception of performance in SE considering gender differences, and indicated their implicit positionality in their threats to validity: *"It is possible that we were biased in interpreting the responses. We avoided this threat as much as possible by having multiple researchers evaluate the data (note that the author team included two women and two men)."* In this example, the authors reflect on how having a gender-balanced research team helped them explore a problem focused on gender differences. By collaboratively evaluating the data, they aimed to minimize gender biases and ensure a more balanced analysis. The first author of [41] (categorized as N/A) was also a co-author in [38], and placed a similar statement in the method section *"Following the methodology of Standpoint Theory, we include a statement of standpoint of the research team. The research team included Black males and females who have strong interests in the disruption of hegemonic normativity in computing."* and also made explicit how they dealt with some of their biases: *"We bracketed our various sets of assumptions by including all variation in the Black women's interviews, not rejecting those experiences that may have been divergent from our own."*. The authors of [42] (action research) although not stating their mixed genders in the text, they bring concerns on *"reflectively questioning our own approach and constantly evaluating it"*. They have perspectives from feminist research and participatory methodologies, aiming to critically engage with their positionality and mitigate biases in their research.

*B. Survey results*

In our analysis, we explored participants' responses to gain a better understanding of their perspectives on three main aspects: (1) how they understand the concept of positionality statement; (2) whether participants or their co-authors have included positionality statements in their own research; and (3) the methods they use to address personal bias in their D&I studies and how this relates to positionality.

**Software Engineering Researchers' Understanding About Positionality Statement**. We grouped the participants' understandings of positionality statements in research into four main themes: self-reflection, researcher influence, contextual awareness, and philosophical stance. Our findings indicate that software engineering researchers see positionality as a way to express how their identities, experiences, and perspectives intersect with their research, as described below:

- **Self-reflection**: Participants described positionality as a means to clarify their personal connection to the research topic. Through this reflection, they consider how their views may shape study design, analysis, and outcomes, not only for transparency but to develop awareness of how personal beliefs and values may influence their work.
- **Researcher influence**: This theme emerged as participants noted that positionality statements allow them to recognize how their lived experiences and social identities, such as cultural or socioeconomic backgrounds, impact their research. By acknowledging these influences, they believe they can address the subjective elements that can shape their interpretations.
- **Contextual awareness**: Participants see positionality statements as a way to place themselves within the study's broader context. Some expressed that this includes articulating how their position relates to the communities under study (e.g., software practitioners or SE students) or the topic under study, informing their engagement approach with the research, and offering readers details about their perspective and potential influences on findings.
- **Philosophical stance**: Some participants view positionality as a reflection of the researcher's worldview and theoretical position. For these participants, a positionality statement offers information regarding their approach to knowledge, including beliefs about objectivity, subjectivity, and the nature of their research questions.

In summary, these four themes —self-reflection, researcher influence, contextual awareness, and philosophical stance— encapsulate the main ways participants understand the purpose and role of positionality statements.

**Positionality Statement Uses in Software Engineering**. In response to how positionality statements have been applied in SE D&I research, we identified four main themes (illustrated in Table II) around the overarching theme of **Positionality Statement Construction**, capturing how participants apply positionality in their research by expressing personal perspectives or acknowledging how their identities and experiences shape their work. The four main approaches identified are:

- **Explicit Placement in Methodology**: Some participants stated that they embed positionality statements within the methodology section. This approach explicitly aligns authors' personal perspective with the research context and providing readers with background on the researcher's connection to the topic.
- **Reflexivity to Address Bias**: Participants described that positionality statements supported them to engage in reflexive practices, considering how personal backgrounds and beliefs might have influenced data interpretation and the study outcomes.
- **Diverse Authorship to Introduce Multiple Perspectives**: Some participants emphasized that they relied on team diversity as a way to include positionality indirectly. By ensuring varied backgrounds within the research team, participants increased their perspectives on the topic, us-ing this to address individual biases. In this scenario, team composition was used as an alternative when individual positionality statements were limited.

Additionally, in our analysis, some participants encountered difficulties incorporating positionality statements into their research due to methodological constraints. For example, double anonymity in review processes can restrict open positioning ( e.g. ,*"It is difficult due to double-anonymity - P15"*), which was a concern we found in HCI literature [14]. Additionally, there was uncertainty among participants about the practical relevance of positionality statements in SE. This finding goes in the opposite direction of other fields that commonly integrate positionality statements explicitly, often using specific approaches to avoid issues [2] with review processes, such as structured sections or carefully framed disclosures that address positionality without compromising anonymity.

In summary, the surveyed SE researchers in D&I apply positionality statements in varied ways, highlighting both the strategies employed and the challenges faced in integrating these statements. Unlike fields such as Social Sciences and HCI, where positionality statements are often explicitly structured, SE researchers tends to adopt a more implicit approach. While other fields dedicate a section for positionality, participants in this study described more subtle methods, embedding positionality through research design choices or team diversity rather than explicitly dedicating sections to it.

TABLE II
USES OF POSITIONALITY STATEMENTS AND SUPPORTING QUOTATIONS

| Use of Positionality Statement | Supporting Quotations |
| --- | --- |
| Explicit Placement in Methodology | P1: "In qualitative research you describe your position explicitly." P7: "We provided a description of the members of the research team, including a statement of their backgrounds, institutions, etc." P8: "I put it in the methodology." |
| Engaging in Reflexivity to Address Bias | P3: "We reflected on what biases we bring to the table and strived to make them explicit, in a concise manner." P11: "A positionality statement is a statement to say how my lived experiences impact the way I conduct research." |
| Incorporating Diverse Authorship to Broaden Perspectives | P6: "Gender-balanced team: starting from having teaching instructors and supervisors of different genders to the ones who are then analysing and writing the paper." P7: "We provided a description of the members of the research team (..)." |
| Challenges in Implementation | P5: "It is difficult due to double-anonymity but we tried." P11: "I can't claim that I have a good answer for how best to do this yet." |

**Research Bias Avoidance in Software Engineering and Its Relation to Positionality Statements**. Regardless of explicitly using positionality statements, we wanted to explore whether bias avoidance practices in SE D&I research were related to positionality, as observed in other fields. Participant P16 had a good perspective by saying that *"there's a deeper issue that positionality statements touch upon, which is that how we approach research from the outset can be influenced by our personal bias"*. Participants' responses indicated also that, while positionality is not always explicitly tied to bias avoidance in SE, many practices align with approaches commonly seen in disciplines that employ positionality statements:

- **Team Diversity and Collaboration**: Researchers built teams with authors from diverse backgrounds, including gender, cultural, and social backgrounds, to integrate multiple perspectives and reduce individual biases. In other fields, positionality statements emphasize acknowledging and incorporating diverse voices to counterbalance researchers' personal views. This collaborative diversity within SE mirrors such positionality practices by embedding a range of perspectives into the research process.
- **Philosophical Reflection**: Researchers engaged in discussions with peers to explore their own methodological choices critically. In fields like social sciences, positionality statements often incorporate reflective analysis of a researcher's stance and assumptions. Similarly, SE researchers practiced philosophical reflection by seeking peer feedback, allowing them to question their biases and support a multi-faceted approach to research design.
- **Self-Reflection**: Researchers employed self-reflective practices to assess how their identities and personal experiences might influence the interpretations of their findings. In other fields, positionality statements involve explicit reflection on a researcher's position to identify and mitigate potential biases. In SE, self-reflection can also serve a similar role by fostering transparency and awareness, helping researchers acknowledge and address personal biases in their analyses.

Together, these practices demonstrate how SE D&I researchers in software engineering have been working to mitigate personal biases. While these approaches may not always be framed explicitly as positionality statements, they reflect core principles of this practice observed in other fields. Table III brings sample codes that lead to these findings.

## V. DISCUSSION

Our analysis of articles of the SLR on Diversity in Software Engineering [20] revealed a significant lack of explicit reflexivity or positionality statements. This scarcity highlights that positionality statements are rarely utilized in empirical qualitative SE research and underscores a critical gap in the usage of reflexivity as a transparent research practice. This confirms such scarcity in our field as recently reported [34]. Even though fundamentally qualitative studies would naturally have such statements Social science studies, the two Ethnography Studies in the SLR did not have that.

Our survey findings show that software engineering researchers understand positionality statements in terms of self-reflection, researcher influence, contextual awareness, and philosophical stance. This is contrasting with the SLR analysis findings, where most articles did not present any traces of reflexivity or positionality. Nevertheless, the findings align in part with the literature, particularly in fields like social sciences, where positionality is seen as essential for research transparency and ethical integrity [8], [11]. In both SE and social sciences, self-reflection and researcher influence are central concepts, acknowledging that researchers' identities and experiences shape their perspectives [17]. This is more implicit in SE, with less emphasis on structured positionality than in social sciences, where researchers openly engage with their own social and cultural positions in a systematic way [22]. Contrastingly, SE researchers see positionality as situational and context-driven, suggesting the potential benefit of more defined frameworks for positionality within the field.

TABLE III
BIAS AVOIDANCE PRACTICES AND SUPPORTING QUOTATIONS

| Bias Avoidance Practice | Supporting Quotations |
|---|---|
| Team Diversity and Collaboration | P2: "Having co-authors from different backgrounds helps to mitigate personal bias." |
| | P5: "It also helps to work in a diverse research team." |
| | P20: "I do this by collaborating and co-authoring with diverse colleagues." |
| Philosophical Reflection | P9: "Discussing methodology with a diverse set of peers." |
| | P19: "Integrating these discussions in articles can ensure clarity in positioning." |
| | P15: "By (...) discussing perspectives, we reduce bias." |
| Self-Reflection | P11: "Positionality statements, like all the other forms of reflection, are tools to avoid biases." |
| | P16: "Describe who I am and what my assumptions are to avoid unwanted influences on research." |
| | P14: "In qualitative research, we try to include reflections on how our positions influence the analysis." |

When comparing positionality practices in SE with those in other disciplines, such as social sciences and human-computer interaction, distinct differences and challenges emerge. Fields like social sciences commonly integrate positionality statements within the methodology section, openly discussing how researchers' backgrounds and beliefs shape the research process [25], [26]. SE researchers, however, face structural constraints like double-anonymity in peer review, which limits transparency around personal perspectives [24]. In fields where reflexivity is essential for understanding power dynamics and researcher-participant relationships, such as social sciences, such constraints are less common, allowing more direct engagement with positionality [21]. The implicit nature of positionality in SE contrasts with that, suggesting that SE could benefit from adopting guidelines that align with its research practices while enabling greater reflexivity. In HCI, such statements help address confirmation bias by acknowledging researchers' perspectives, increasing transparency, and minimizing unintended influence on data collection and interpretation through strategies (e.g., multiple coders, agreement processes) aiming to reduce biases of a particular researcher [43].

Finally, SE researchers employ practices to avoid personal bias that resemble positionality statements in other fields, though often without the explicit terminology. Collaborative research with diverse teams, philosophical reflection, and self-reflection mirror positionality practices in disciplines that emphasize addressing normative biases and ensuring inclusive perspectives [5], [23]. Although SE researchers rarely frame these practices as positionality, they are engaging in reflexive approaches that fulfill similar goals—acknowledging and mitigating personal biases. The approach suggests that SE researchers already value reflexivity, yet the field can benefit from clear guidelines to formally incorporate positionality statements, fostering a research environment that recognizes and values diverse perspectives. The quality of SE qualitative research can improve if researchers emphasize reflexivity [34].

## VI. IMPLICATIONS

The importance of reflexivity in SE was recently reported [34], suggesting that fostering a culture of openness and authenticity within research teams and the broader academic community is vital for effectively implementing reflexivity in qualitative SE research. Our research is the first to analyze the use of positionality statements in our field, bringing attention to a practice that has long been established in more mature disciplines like social sciences and can be used as an initial step to educate researchers toward reflexivity practices. In these disciplines, positionality statements have become a structured research practice, developed over decades to promote transparency, reflexivity, and ethical rigor [8], [21]. As empirical SE continues to evolve, it has the opportunity to draw from these established practices, such as adopting positionality statements to enhance its methodological depth. This is not a matter of having a check-list item for a shallow statement, but incorporating these reflective practices that can help software engineering better address the complex human aspects that are becoming increasingly central to the field.

This approach is particularly relevant for studies on equity, diversity, and inclusion, as well as research involving hidden or underrepresented populations in software engineering. In such work, sensitivity to a range of viewpoints is essential to accurately represent the experiences and challenges of diverse participant groups. Positionality statements offer a structured way for researchers to acknowledge their perspectives, which can help avoid blind spots and build trust with study participants [24], providing a means to foster inclusivity and ensure that researchers approach these communities thoughtfully, enhancing the reliability and applicability of findings.

Moreover, integrating positionality into established sections of software engineering papers, such as threats to validity, could help researchers document the ways their backgrounds influence their interpretations. This approach would enhance validity considerations, particularly in studies with a strong human element (e.g., case studies and ethnographies), where researcher perspectives often shape the framing and outcomes of research. Thus, by adopting positionality as a component of research validity, software engineering can build a more reflective and transparent research culture.

Finally, software engineering research has the opportunity to embrace positionality statements in a way that reflects its distinct intersection of human and technological considerations.

Unlike social sciences, where positionality is often linked to social identities and relationships, software engineering could expand the concept to include professional and technical experiences, under a socio-technical lenses. For example, researchers who were once practitioners in software testing or enthusiasts in open-source software could incorporate their professional experiences as part of their positionality. By integrating both personal and professional dimensions, SE research could create a nuanced approach to positionality, incorporating both human and technical influences on research perspectives, ultimately enriching the relevance of findings.

## VII. Lessons Learned and Recommendations

Based on participants' narratives, a few key lessons emerged. Positionality statements in software engineering are often applied implicitly, with researchers acknowledging that personal and professional experiences shape their perspectives but lacking structured guidance for formally documenting these influences. Institutional factors (e.g., double-anonymity in peer review, SE's traditional focus on objectivity), bring challenges to explicitly adopting positionality statements.

Given SE's intersection of human and technical considerations, there is an opportunity to develop approaches to positionality addressing both dimensions. By fostering a more reflexive and inclusive research culture, SE can advance its understanding of human factors while enhancing methodological rigor. To support the integration of positionality in SE research, we propose the following recommendations:

**Normalize Positionality Statements in SE**: Empirical SE research could encourage positionality statements as a standard practice, particularly in sections like method or threats-to-validity, where researchers can clarify how their backgrounds and experiences might have shaped study outcomes.

**Engage in Reflective Practice at the Start of Research**: Researchers could start the project by reflecting on their identities, backgrounds, and experiences, considering how these factors may influence their questions, methods, and findings.

**Promote Diverse and Collaborative Research Teams**: Empirical SE research guidelines could motivate researchers to foster team diversity across gender, culture, and professional expertise to reduce individual biases, enhance inclusivity, and tackle complex software engineering challenges requiring multi-faceted views on societal impact.

**Expand Positionality Statements to Include Professional Backgrounds**: Researchers in software engineering could incorporate technical expertise in their positionality statements by including relevant professional experiences that can help acknowledge how their professional background influences their research motivations and perspectives.

**Create SE-Specific Guidelines for Positionality**: Empirical software engineering research could develop specific guidelines to provide SE researchers with clear directions on structuring and integrating positionality tailored to the SE context.

## VIII. Threats to Validity

This study focused exclusively on authors of papers related to Diversity and Inclusion (D&I) in Software Engineering. While this population was selected to gather perspectives from researchers likely familiar with positionality and its potential impact, it limits the scope of our findings. Future research should expand to include authors investigating a wider range of software engineering topics to develop a broader perspective of positionality practices across the field.

Additionally, purposive sampling was employed to reach a targeted group. While this approach was suitable for gathering relevant perspectives, it proved limited as we were only able to collect data from 21 participants, out of 285 valid messages. There is a potential bias of only invitees aware of the positionality concept being motivated to answer the survey, which is in accordance with the low occurrence of positionality discussion in articles in the SLR analysis. Nevertheless, this limitation suggests that additional sampling methods, such as snowball sampling, could be integrated into future studies to enhance the diversity and volume of responses, potentially leading to richer data and more nuanced outcomes.

Finally, this study does not aim for generalization. First, we received fewer responses than initially anticipated, which limits the representativeness of our sample. As a qualitative study, the goal is not to generalize findings to the broader field but rather to provide in-depth, context-rich discussions. Thus, we expect that our findings could be analyzed analytically, offering valuable information to inform and guide future research on employing positionality in SE research.

## IX. Conclusions

This study focused on positionality statements, a practice well-established in human-centered disciplines, where it serves to enhance research transparency, reflexivity, and ethical rigor. Positionality statements involve researchers openly acknowledging how their identities, backgrounds, and perspectives may shape their research, an approach particularly valuable in areas addressing complex human and societal issues. While SE traditionally centers on technical development, it shares with these fields a focus on human dynamics, e.g., in software development, making positionality a timely and relevant topic. Our findings indicate that SE researchers are beginning to understand and apply positionality in ways that reflect an awareness of personal influences on research. Participants generally view positionality as a means for self-reflection, contextual awareness, and recognition of how their personal and professional backgrounds may shape interpretations. Currently, there is emerging understanding on the value of positionality in SE research, but its usage vary, e.g., some researchers embedding positionality statements explicitly, others approaching it implicitly through team diversity or self-reflection.

In summary, our findings demonstrate that incorporating positionality can enrich software engineering research by fostering inclusivity, enhancing ethical rigor, and promoting a deeper understanding of how personal and professional perspectives shape outcomes. As the field continues to address complex social and technical issues, adopting positionality as a reflective practice can help build a more thoughtful, transparent, and inclusive research culture within SE.